\documentstyle[12pt]{article}

\begin{document}
\begin{titlepage}
\begin{center}

\today    \hfill       MIT-CTP-2266
\hskip .5truein       \hskip .5truein hep-th/9401041\\

\vskip .2in

{{\Large \bf Strings and Two-dimensional QCD for Finite
N\footnote{This work was supported in part by the divisions of Applied
Mathematics of the U.S. Department of Energy under contracts
DE-FG02-88ER25065 and DE-FG02-88ER25066 and in part by the National
Science Foundation under grant PHY90-21984.}}}

\vskip .15in

John Baez\footnote{\tt baez@math.ucr.edu}\\
{\small {\em
 Department of Mathematics \\
	University of California  \\
	Riverside CA 92521, USA}}

\vskip .15in

Washington Taylor\footnote{\tt wati@mit.edu}\\

{\small {\em
Center for Theoretical Physics \\
Laboratory for Nuclear Science \\
and Department of Physics\\
Massachusetts Institute of Technology \\
Cambridge, Massachusetts 02139, USA}}
\end{center}
\vskip .15in

\begin{abstract}
The string theory description of $SU(N)$ Yang-Mills on an arbitrary
two-dimensional manifold, previously developed for the large $N$
asymptotic expansion, is extended to include finite values of $N$.
The theory is considered from two points of view, first using a
canonical Hamiltonian formulation, second using a global description
of the partition function.  In both formalisms, the effect on the
string theory of taking a finite value of $N$ is described by a local
projection operator which has a simple description in terms of the
symmetric group $S_n$.
\end{abstract}
\end{titlepage}
\newpage
\renewcommand{\thepage}{\arabic{page}}
\setcounter{page}{1}

\section {Introduction}
\setcounter{equation}{0}
\baselineskip 18.5pt

Recently, there has been a renewed interest in Yang-Mills theory in
two dimensions.  The explicit description of the
asymptotic $1/N$ expansion of the partition function in terms of an
equivalent string theory gives hope that a similar connection might be
found for four-dimensional gauge theories, in particular for
four-dimensional QCD.  A review of the recent work in this area is
given in \cite{gt3}.  In this paper we extend the string description
of the $SU(N)$ gauge theory to obtain an exact string-theoretic
description for arbitrary finite $N$.

It was first shown by Migdal \cite{migdal} almost 20 years ago that in
two dimensions, the pure Yang-Mills gauge theory with any gauge group
is exactly soluble, and can be described by a triangulation-invariant
lattice theory.  This work was later expanded upon
\cite{rusakov,witten,fine,BT} to describe the pure gauge
theory on an arbitrary Riemann surface.  The general formula for the
partition function of the two-dimensional Yang-Mills theory on a
Riemann surface ${\cal M}$ of genus $G$ and area $A$ is given by
\begin{eqnarray}
Z(G,\lambda A,N) & = &  \int[{\cal D} A^\mu]
{\rm e}^{- {1\over 4 {\tilde g}^2} \int_{\cal M}
 {\rm Tr} (F \wedge \star F)}  \nonumber\\
  & = & \sum_{R} (\dim R)^{2 - 2G}
  {\rm e}^{-\frac{\lambda A}{2 N}C_2(R)}, \label{eq:partition}
\end{eqnarray}
where the sum is taken over all irreducible representations of the
gauge group, with $\dim R$ and $C_2(R)$ being the dimension and
quadratic Casimir of the representation $R$. The coupling constant
$\lambda$ is related to the gauge coupling $\tilde{g}$ by $\lambda =
\tilde{g}^2 N$.

It was recently shown \cite{gross,minahan,gt1,gt2} that by expanding
the sum over representations in (\ref{eq:partition}) in powers of the
gauge group parameter $N$, the asymptotic expansion for the
partition function as $N \to \infty$
can be written to all orders in $1/N$ in terms of a sum
over maps from a string world sheet onto the manifold ${\cal M}$.
The partition function given by this asymptotic expansion can be
regarded as a theory in its own right, the ``large $N$ theory.''
The string maps summed over are maps that are
singular only at a finite number of points.  The partition function
(\ref{eq:partition}) can then be rewritten in the form
\begin{equation}
Z (G, \lambda A, N) = \int_{\Sigma({\cal M})} \,{\rm d} \nu\, W (\nu),
\end{equation}
where $\Sigma ({\cal M})$ is a set of branched covering maps
\begin{equation}
\nu: {\cal N} \rightarrow {\cal M},
\end{equation}
{}from Riemann surfaces ${\cal N}$ of arbitrary genus $g$ onto ${\cal M}$,
and the weight of each map in the partition function is given by
\begin{equation}
W (\nu) =   \pm
\frac{N^{2-2 g}}{ | S_\nu |}
{\rm e}^{- \frac{n \lambda A}{2}}.
\label{eq:mapwt}
\end{equation}
In the weight, $n$ is the degree of the map $\nu$ and $| S_\nu |$ is
the symmetry factor of the map (the number of diffeomorphisms $\pi$ of
${\cal N}$ which satisfy $\nu \pi = \nu$).  The sign of the weight
depends upon the set of singular points in the map $\nu$.  The types
of singularities allowed for maps $\nu$ in the set $\Sigma ({\cal M})$
depend upon the gauge group and the genus $G$.  In \cite{gt2}, it was
shown that in the large $N$ expansion, the vacuum expectation values
of Wilson loops can also be expressed in terms of a sum over open
string maps, where the boundary of the string world sheet is mapped to
the Wilson loop.

In the initial work \cite{gt1,gt2}, the large $N$ expansion was
studied for the gauge groups $SU(N)$ and $U(N)$.  The $SU(N)$ theory
describes two-dimensional QCD (QCD${}_2$).  Recently, this work has
been expanded to the other gauge groups $SO(N)$ and $Sp(N)$
\cite{nrs,r}.
For these gauge groups, the string world sheet may not be orientable;
the exponent of $N$ in (\ref{eq:mapwt}) is then simply the Euler
characteristic of the string world sheet.
In other recent work, connections have been made
between QCD${}_2$ and several other theories of current interest
\cite{douglas,mp,cdmp}.  The theory has also been studied on the
lattice, and shown to give rise to an equivalent theory of
triangulated strings \cite{kostov}.  The behavior of the partition
function of the $U(N)$ theory on the sphere has been studied by
Douglas and Kazakov
\cite{dk}.  They showed that for small areas $\lambda A < \sqrt{\pi}$,
the partition function in the large $N$ limit undergoes a phase
transition, and that in the small area phase the partition function
becomes trivial.  Such a phase transition does not occur on higher
genus Riemann surfaces, and does not occur for finite values of $N$.
Recently, using similar methods the vacuum expectation value for a
simple Wilson loop on the sphere has been studied and shown to behave
similarly to the partition function \cite{bo,rusakov3,dak,mp3}.

In this paper, we extend the earlier results to describe a string
theory for finite $N$ Yang-Mills on an arbitrary two-dimensional
manifold ${\cal M}$.  We restrict attention in this paper to $SU(N)$,
although the results are easily extended to other gauge groups.  The
primary result is that this theory can also be described in terms of a
weighted sum over string maps; however, in the finite $N$ theory, it
is necessary to introduce an additional projection operator into the
string theory picture.  Although this projection operator complicates
the string theory somewhat, one of the other complicating features of
the large $N$ theory disappears for finite $N$.  Specifically, in the
string theory description of large $N$ QCD${}_2$, it is necessary to
consider two ``chiral'' sectors, corresponding to strings with two
distinct orientations.  When the string partition function is written
in terms of a sum over maps from a string world sheet to the target
space ${\cal M}$, the existence of these two sectors makes it
necessary to give an orientation to the string world sheet and to
distinguish between orientation-preserving and orientation-reversing
maps.  In the finite $N$ theory, only one chiral sector need be
counted, so that the resulting string theory is essentially a theory
of unoriented strings.

There are several possible ways to view the projection operator
for the finite $N$ QCD${}_2$ string theory.  We explore two such
views in this paper.  First, we analyze the partition function of
the finite $N$ theory in the Hamiltonian formulation.
In the Hamiltonian picture, the Hilbert space of QCD${}_2$ on a slice
$S^1$ of ${\cal M}$ is simply the set of class functions on the gauge
group $SU(N)$.  Two natural sets of states spanning this Hilbert space
are the set of characters of irreducible representations $\chi_{R}
(U)$, and the set of products of traces
\begin{equation} \Upsilon_{\sigma} (U) = \prod_{j = 1}^{s} ({\rm
Tr}\; U^{k_j}), \label{eq:stringstates} \end{equation}
where the matrices $U$ are in the fundamental representation, and
$\sigma =\{k_1, \ldots, k_s\}$ is a set of positive integers with $0 <
k_1 \leq \cdots \leq k_s$.
The class functions given by products of traces can be interpreted in
terms of a set of $s$ strings winding around the circle $k_j$ times;
we will refer to these functions as ``string'' states.  As $N \to
\infty$, any finite set of distinct string states become linearly
independent.  For fixed finite $N$, however, the string states have
linear dependencies, given by the Mandelstam identities
\cite{GamTri}.  To develop a string-theoretic interpretation of the
finite $N$ gauge theory it is desirable to use the string states.  The
redundancy of these states can be eliminated by considering an
``extended'' state space consisting of finite linear combinations of
{\it formal} string states corresponding to all the $\Upsilon_\sigma$
functions.  As described in \cite{Baez}, the ``physical'' Hilbert
space is then the quotient of this extended state space by the
Mandelstam identities, suitably completed to form a Hilbert space.  In
this paper we follow an alternative method.  First we complete the
extended state space to obtain a Hilbert space which has the string
states as an orthogonal basis, and then we identify the physical
Hilbert space with a subspace.  We explicitly construct the operator
on the extended Hilbert space that projects down to this subspace;
this projection operator has a simple form in terms of the string
basis.

There is a natural Hamiltonian on the extended Hilbert space,
which is most simply expressed in terms of the character basis.
In fact, the Hamiltonian operator is diagonal in this basis and
acts on the representation $R$ by multiplication by $C_2 (R)$ times a
constant.  In terms of the string basis, the Hamiltonian is no longer
diagonal but contains interaction terms which describe the change in
string windings caused by singular points such as branch points and
infinitesimal tubes \cite{jevicki,mp}.  In the finite $N$ theory, we can use
precisely the same Hamiltonian, restricted to the subspace of
physical states.

The second view of the projection operator in the finite $N$ theory
arises when we analyze the partition function in a purely
group-theoretic fashion, using techniques analogous to those presented
in \cite{gt1,gt2}.  These previous papers dealt with the large $N$
expansion.  If one replaces the sum over irreducible representations
in (\ref{eq:partition}) by a sum over all Young diagrams with finitely
many boxes one obtains the partition function for the large $N$ theory
(or, more precisely, a single ``chiral sector'' thereof).  Using the
theory of Young diagrams this sum can be rewritten as a sum over
elements of the symmetric groups $S_n$.  The group-theoretic terms in
this expression can all be grouped into a single $\delta$ function of
a product of elements of $S_n$.  This $\delta$ function can then be
interpreted as a topological condition on permutations of sheets of a
covering space, where a permutation $\sigma \in S_n$ at a point $x \in
{\cal M}$ describes a multiple branch point singularity in the
covering map; the positions of the singularities are integrated over,
giving factors of the area of ${\cal M}$ in the partition function.
This gives a string-theoretic interpretation of the partition function
for large $N$ $QCD$ in two dimensions.

Irreducible representations of $SU(N)$ correspond to Young diagrams
with $< N$ rows.  In the finite $N$ case, we find that the restriction
to irreducible representations of $SU(N)$ in the sum
(\ref{eq:partition}) changes the $\delta$ function of the group
theoretic terms in the partition function in a way that corresponds to
the insertion of a projection operator at a single fixed but arbitrary
point on the manifold ${\cal M}$.  We find that this projection
operator is mathematically equivalent to the projection operator
constructed in the Hilbert space picture, when both are expressed in
terms of the symmetric group.  The ``projection point'' plays a role
similar to that of the $\Omega$-points which arise in the large $N$
theory on higher genus Riemann surfaces \cite{gt1,gt2}.
($\Omega$-points also appear in the finite $N$ theory when ${\cal M}$
is not a torus.)  The projection points, like the $\Omega$-points, are
points at which the string maps are allowed to have extra
singularities.  Just as the $\Omega$-points probably indicate some
global structure in the string theory (as opposed to local
interactions, which correspond to integrals over $\cal M$), the
projection point will hopefully give some insight into what features
must be added to a string action formalism for finite $N$ QCD${}_2$.

In Section 2, we review the Hamiltonian description of QCD${}_2$ and
the string interpretation of this theory in terms of bosonic creation
and annihilation operators.  Within this framework, we explicitly
construct the projection operator from the extended state space to
the physical Hilbert space and describe the resulting string picture
of the partition function for finite N QCD${}_2$ on the torus.  We
observe that for $SU(2)$, the projection operator has a particularly
simple form.  In Section 3, we rederive the projection operator of
Section 2 by considering the partition function of the finite $N$
gauge theory from a purely group-theoretic point of view.  We expand
(\ref{eq:partition}) as a formal power series and show that the
restriction to irreducible representations of $SU(N)$ can be described
algebraically by the insertion of a projection operator which is
mathematically equivalent to the one derived in Section 2.  Finally,
in Section 4 we review our results and discuss the possible
application of this work to Euclidean quantum gravity in 3 dimensions.

\section {Hilbert Space Picture}
\setcounter{equation}{0}
\baselineskip 18.5pt
\renewcommand{\tilde}{\widetilde}

We will now proceed to describe the string theory of QCD${}_2$ from
the Hamiltonian point of view.  We begin by considering the theory on
the cylinder $S^1
\times [0,1]$.  For a fixed finite value of $N$,
the Hilbert space ${\cal H}_N$ of QCD${}_2$ associated
with a spatial $S^1$ slice is simply the linear space of class
functions on the gauge group $SU(N)$.  As described in Section 1,
there are two natural bases for this Hilbert space.  One basis is the
set of characters $\chi_R (U)$ of irreducible representations of
$SU(N)$.  This basis is orthonormal under the inner product
\begin{equation}
\int{\rm d} U \chi_R (U) \chi_{R'} (U^{\dagger}) = \delta_{R,R'},
\label{eq:circle}
\end{equation}
where ${\rm d} U$ is the invariant Haar measure on $SU(N)$ normalized to
$\int {\rm d} U= 1$.  Another basis for ${\cal H}_N$ is the set of
string functions $\Upsilon_\sigma$ defined by (\ref{eq:stringstates}).
For each $n$, there is a set of string functions   indexed by
(conjugacy classes of) permutations $\sigma \in S_n$.
For finite $N$, the set of string  functions spans ${\cal H}_N$, but
also contains linear dependencies.  For $N > n$, the set of string
functions $\Upsilon_\sigma$ with $\sigma \in S_n$ is linearly
independent; these class functions are related to the irreducible
characters of $SU(N)$ by the Frobenius relations
\begin{eqnarray}
\chi_R (U) & = & \sum_{\sigma\in S_n}\frac{\chi_{R}(\sigma)}{n!}
\Upsilon_{\sigma} (U),  \label{eq:frobenius} \\
\Upsilon_{\sigma}(U) & = &  \sum_{R\in Y_n} \chi_{R} (\sigma)
\chi_{R} (U),
\end{eqnarray}
where $\chi_R (U)$ is the character of $U$ in the representation $R$
of $SU(N)$ associated with a Young diagram containing $n$ boxes, and
$\chi_R (\sigma)$ is the character of the permutation $\sigma$ in the
representation of $S_n$ associated with the same Young diagram. (We
denote by $Y_n$ the set of all Young diagrams with $n$ boxes.)

Rather than using the actual Hilbert space associated with irreducible
representations of $SU(N)$ for finite $N$, we will find it more
convenient to consider an extended Hilbert space $\tilde{\cal
H}$, as in \cite{Baez}.  This Hilbert space has an orthonormal
basis given by the set of all Young diagrams with a
finite number of boxes; we write $| R \rangle$
for the state corresponding to the Young diagram $R$.
Given a permutation $\sigma \in S_n$, we then define the ``string
state'' $|\sigma\rangle$ in $\tilde{\cal H}$ by the Frobenius relation
\begin{equation}
| \sigma \rangle  =   \sum_{R\in Y_n} \chi_{R} (\sigma)
| R \rangle.
\label{eq:frobenius2} \end{equation}
If the permutation $\sigma$ has cycles of length $k_1, \dots,
k_s$, the state $|\sigma\rangle$ corresponds physically to a
state in which there are $s$ strings with winding numbers $k_1,
\dots, k_s$.  Note that the string states corresponding to
conjugate elements in $S_n$ are equivalent, so that
\begin{equation}
 | \sigma \rangle = | \tau \sigma \tau^{-1} \rangle  \; \; \;
\forall \sigma, \tau \in S_n.
\end{equation}

One can show that the other Frobenius relation,
\begin{equation}
| R \rangle =  \sum_{\sigma\in S_n}\frac{\chi_{R}(\sigma)}{n!}
| \sigma \rangle,
\end{equation}
follows from this definition.  Moreover, one has the following
formula for the inner product of string states
\begin{equation}
\langle \sigma | \tau \rangle =
\frac{\delta_{T_\sigma,T_\tau} n!}{| T_\sigma |},
\end{equation}
where $T_\sigma$ is the conjugacy class of elements in
$S_n$ that contains $\sigma$.
We can therefore rewrite the identity operator in $\tilde{\cal H}$
as
\begin{equation}
1 = \sum_{n} \frac{1}{n!}  \sum_{\sigma \in S_n}
| \sigma \rangle \langle \sigma |.
\end{equation}

For a fixed finite value of $N$, the irreducible representations
of $SU(N)$ are in one-to-one correspondence with the Young
diagrams having $< N$ rows.  Thus
the Hilbert space ${\cal H}_N$ for the finite $N$ theory may be
identified with the subspace of $\tilde{\cal H}$ spanned by states
$|R\rangle$ corresponding to Young diagrams with $< N$ rows.
We call this the ``physical'' subspace of $\tilde{\cal H}$.  Note that
the inner product on ${\cal H}_N$ coincides with the inner
product on the physical subspace inherited from that on
$\tilde{\cal H}$.
In terms of the string basis, the
projection operator onto the physical subspace is written
\begin{eqnarray}
P_N & =  &\sum_{n} \sum_{R \in Y^{N}_n}| R \rangle \langle R|
\nonumber \\
&= &\sum_{n} \sum_{R \in Y^{N}_n}
\sum_{\sigma, \tau \in S_n}
\frac{1}{n!^2} | \sigma \rangle \langle \sigma| R \rangle
\langle R|\tau \rangle \langle \tau |  \label{eq:projection2}\\
& = & \sum_{n} \sum_{R \in Y^{N}_n}
\sum_{\sigma, \tau \in S_n}
\frac{d_R}{n!^2} \chi_R (\sigma \tau^{-1})
| \sigma \rangle \langle \tau |, \nonumber
\end{eqnarray}
where $Y^{N}_n$ is the set of Young diagrams in $Y_n$ with fewer than
$N$ boxes in each column, and $d_R = \chi_R (1)$ is the dimension of
the representation $R$ of $S_n$.

We will now discuss the Hamiltonian on the physical and extended
Hilbert spaces ${\cal H}_N$ and $\tilde{\cal H}$.  We assume that the
manifold ${\cal M}$ on which we are describing the theory (in this
case the cylinder) is equipped with a metric $g$, such that $L$ is the
length of the slice $S^1$ associated with the Hilbert space, and such
that $g_{tt}= 1$.
In terms of the basis of characters $\chi_R (U)$
for the physical Hilbert space, the Hamiltonian $H$ is a diagonal
operator which acts by multiplication by $\lambda L C_2 (R)/2 N$,
where $C_2 (R)$ is the quadratic Casimir of the representation $R$.
(See for example \cite{douglas}.)

For a representation $R$ associated with a Young diagram with $n$
boxes and with a set of symmetric group characters $\chi_R (\sigma),$
$\sigma \in S_n$, the $SU(N) $ quadratic Casimir is given by
\begin{equation}
C_2 (R) = n N + \frac{n (n-1)\chi_{R} (T_2)}{d_R} - \frac{n^2}{N},
\label{eq:casimir}
\end{equation}
where $T_2$ is the conjugacy class of elements in $S_n$ containing one
cycle of length 2 and $n - 2$ cycles of length 1.
We can thus define an associated Hamiltonian $\tilde{H}$ on the
extended Hilbert space $\tilde{\cal H}$ by
\begin{equation}
\tilde{H} | R \rangle = \frac{\lambda L}{2}
\left(n + \frac{n (n-1)\chi_{R}
(T_2)}{N d_R} - \frac{n^2}{N^2}\right) | R \rangle,
\end{equation}
where $n$ is the number of boxes in the Young diagram $R$.  From
(\ref{eq:casimir}), it is clear that $\tilde{H}$ agrees with $H$ on
the physical Hilbert space.

The Hamiltonian $\tilde{H}$ has a simple description in terms of
bosonic raising and lowering operators \cite{jevicki,mp}.  In the string
basis, the extended Hilbert space $\tilde{\cal H}$ is equivalent to a
bosonic Fock space, where the ground state $| 0 \rangle$ corresponds
to a state with no string excitations, and the bosonic operators
$a_k,a_k^{\dagger}$ satisfying $[a_k,a_l^{\dagger}] = k
\delta_{k,l}$ correspond to the annihilation and creation respectively
of a string winding $k > 0$ times around the circle.  Note that we are
only including strings winding a positive number of times around the
circle; this corresponds to the fact that we are constructing a theory
of unoriented strings.  In terms of these operators, the Hamiltonian
on the extended Hilbert space can be written
\begin{equation}
\tilde{H} = \frac{\lambda L}{2}
\left[ \sum_{n > 0} a_n^{\dagger} a_n
+ \frac{2}{N} \sum_{n,m > 0}
(a_{n + m}^{\dagger} a_n a_m + a_n^{\dagger} a_m^{\dagger} a_{n + m})
- \frac{1}{N^2}  (\sum_{n > 0} a_n^{\dagger} a_n)^2
\right].
\end{equation}
It is convenient for the string interpretation to write this
Hamiltonian as a sum of terms
\begin{equation}
\tilde{H} = H_0 + \frac{1}{N}  H_1  - \frac{1}{N^2}  (H_h + H_t),
\end{equation}
where
\begin{eqnarray}
H_0 & = & H_h =\frac{\lambda L}{2}  \sum_{n > 0} a_n^{\dagger} a_n\\
H_1 & = & \lambda L \sum_{n,m > 0}
(a_{n + m}^{\dagger} a_n a_m + a_n^{\dagger} a_m^{\dagger} a_{n +
m})\\
H_t & = & \frac{\lambda L}{2}\left[
(\sum_{n > 0} a_n^{\dagger} a_n)^2 -\sum_{n > 0}
a_n^{\dagger} a_n\right].
\end{eqnarray}

To interpret these terms string-theoretically we first consider the
partition function for a torus.  On the torus of area $A$ with cycles
of length $L$ and $\beta$, the partition function for the finite $N$
theory is given by
\begin{equation}
Z (1, \lambda A, N) = {\rm Tr}_{\tilde{{\cal H}}}\; P_N \;{\rm e}^{- \beta
\tilde{H}}.
\label{eq:finitepartition}
\end{equation}
Note that dropping the projection operator $P_N$ from this expression
gives precisely the partition function for a single chiral sector of
the asymptotic expansion of the full theory for large $N$, as
described in \cite{gt1,gt2}.  We can now describe this partition
function in terms of a sum over string maps with the types of
singularities familiar from the large $N$ theory; the projection
operator $P_N$ gives rise to a single additional singular point in the
string covering map.  The leading term in the Hamiltonian, $H_0$, can
be interpreted as the ``free'' string Hamiltonian.  This term gives a
contribution of ${\rm e}^{-\frac{n \lambda A}{2} }$ for a string
configuration of total winding number $n$ on a manifold of area $A$.
By expanding the exponential of the remaining terms in the Hamiltonian
in (\ref{eq:finitepartition}), we can interpret the interaction terms
$H_1, H_t,H_h$ as describing singularities associated with
interactions between strings.  For instance, the interaction term
$H_1$ precisely describes the string interaction due to a simple
branch point in the map from the string world sheet onto the target
space.  At such a branch point, either two strings with winding
numbers $k,l$ combine to form a single string with winding number $k +
l$, or a single string with winding $k + l$ splits to form 2 strings
with winding numbers $k$ and $l$.  An example of such a branch point
interaction is shown in Figure~\ref{f:branchpoint}.  {\small
\begin{figure}
\centering
\begin{picture}(200,100)(- 100,- 50)
\thicklines
\put(- 50, 0){\oval( 10, 45)}
\put( 50, 0){\oval( 10, 45)[r]}
\put(50,  17.5){\oval( 10, 10)[t]}
\put(50, -17.5){\oval( 10, 10)[b]}
\put(- 50, 0){\oval(  15, 50)}
\put( 50, 0){\oval(  15, 50)[r]}
\put( 50,  17.5){\oval(  15, 15)[t]}
\put( 50, - 17.5){\oval(  15, 15)[b]}
\put( 42.5,17.5){\line( 0, -1){ 10}}
\put( 45,17.5){\line( 0, -1){ 10}}
\put( 42.5,-17.5){\line( 0, 1){ 10}}
\put( 45,-17.5){\line( 0, 1){ 10}}
\put(- 50, 25){\line(1,0){100}}
\put(- 50, -25){\line(1,0){100}}
\put(42.5,7.5){\line(1, - 6){2.5}}
\put(42.5,-7.5){\line(1, 6){1}}
\put(45,7.5){\line( -1,- 6){ -1}}
\thinlines
\multiput(0,0)(6, 0){ 7}{\line(1,0){3}}
\put(0,0){\circle*{5}}
\put(- 50, - 35){\makebox(0,0){ $|1,1 \rangle$}}
\put( 50, -35){\makebox(0,0){ $|2 \rangle$}}
\put(0,  10){\makebox(0,0){ $a_2^{\dagger} a_1^2$}}
\end{picture}
\caption[x]{\footnotesize Branch point singularity associated with $H_1$}
\label{f:branchpoint}
\end{figure}
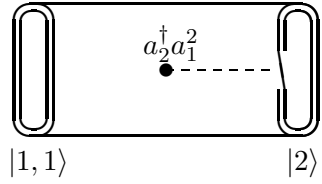}
The factor of $1/N$ associated with this interaction term is precisely
the factor which one expects for a singularity which decreases the
Euler characteristic of the string world sheet by 1.  Similarly, the
remaining interaction terms $H_t$ and $H_h$ give the contributions
{}from adding a string-string interaction between any pair of strings
such as would arise from an infinitesimal tube connecting the sheets
of the string world sheet, and a self-interaction term coming from the
addition of infinitesimal handles to the string world sheet.  In this
way, the partition function can be rewritten as a sum over possible
covering maps associated with sequences of string interactions located
at arbitrary points on the target space ${\cal M}$.  Because the
interaction terms carry factors of $\lambda A$, this gives a natural
measure to the space $\bar{\Sigma} ({\cal M})$ of covering spaces,
where each singularity is given an integration measure proportional to
the area measure on ${\cal M}$.

The only feature which remains to be interpreted in the string picture
is the projection operator $P_N$.  This operator can be simply
interpreted in the same string language as being associated with an
extra singularity point where a nontrivial string interaction occurs.
Because this singularity point does not carry a factor of the target
space area, we will fix a point $p\in {\cal M}$ where this singularity
occurs.  From (\ref{eq:projection2}), we see that a singularity
associated with a permutation $\sigma\in S_n$ of sheets on the string
covering space must give an extra factor of
\begin{equation}
P^{(N)} (\sigma) = \sum_{R \in Y^{N}_n} \frac{d_R}{n!} N^{n - K_\sigma}
\chi_R (\sigma),
\end{equation}
where $K_\sigma$ is the number of cycles in the permutation $\sigma$.
The factor of $N^{n - K_\sigma}$ arises because of the change in the
Euler characteristic of the covering space due to the permutation
$\sigma$.

Thus, we have derived a string picture for the finite $N$ $SU(N)$
gauge theory on the torus.  In this theory, the set of maps
$\bar{\Sigma}({\cal M})$ is equal to the set of all topologically
distinct covering maps $\nu$ from a surface ${\cal N}$ of arbitrary
genus $g$ onto ${\cal M}$, with an arbitrary number of branch point,
tube, and handle singularities, the positions of which are a set of
modular parameters for $\nu$ each carrying a measure equal to $\lambda
{\rm d} A$.  Because the Hilbert space only contains strings with
positive winding, we need not consider ${\cal N}$ to have a fixed
orientation; however, the manifold ${\cal N}$ must be orientable,
since an orientation on ${\cal M}$ can always be pulled back to ${\cal
N}$.  The maps in $\bar{\Sigma} ({\cal M})$ are allowed to have an
additional singularity at the fixed point $p \in {\cal M}$ associated
with an arbitrary permutation $\sigma$ in the sheets of the covering
space.  Such a singularity carries a weight factor of $P^{(N)}
(\sigma)$.  By a careful analysis of the combinatorics, one can verify
that the constant factor associated with a given cover $\nu$ is
precisely $1/| S_\nu |$.  Thus, we find that the partition function
for the finite $N$ $SU(N)$ gauge theory is given by
\begin{equation}
Z (1, \lambda A, N) = \int_{\bar{\Sigma}({\cal M})} {\rm d} \nu W (\nu),
\end{equation}
with the weight of each map in the partition function being given by
\begin{equation}
W (\nu) =    (-1)^i P^{(N)}  (\sigma)
\frac{N^{2-2 g}}{ | S_\nu |}
{\rm e}^{- \frac{n \lambda A}{2}},
\end{equation}
where $i$ is the number of branch points in the map $\nu$, and
$\sigma$ is the permutation associated with the singularity at the
point $p$.

With the exception of the singularity at $p$, which we will refer to
as a ``projection point'' singularity, this is precisely the formula
for the asymptotic expansion of the partition function in a single
chiral sector of the large $N$ theory.  The projection point, which
exists at a fixed point in ${\cal M}$ and does not carry a factor of
the area, is analogous to the $\Omega$-points which arise in the large
$N$ theory on higher genus Riemann surfaces.

As a particularly simple example of the finite $N$ theory,  we
consider the case $N = 2$.  In this case, for any value of $n$, there
is only a single Young diagram in the set $Y^{2}_n$.  The
corresponding representations of the symmetric group are the
completely symmetric representations, which are trivial and have
$\chi_R (\sigma)= 1$ for all $\sigma$.  Thus, in this case the weight
of any permutation $\sigma$ at a projection point simplifies to
\begin{equation}
P^{(2)}  (\sigma) =\frac{2^{n-K_\sigma}}{ n!}.
\end{equation}

It is fairly straightforward to generalize the Hilbert space formalism
used in this section to  describe the finite $N$ theory on higher
genus Riemann surfaces.     Along the lines of the axiomatic approach
to conformal field theory \cite{Segal}, one can introduce a state
\begin{equation}
D \in \tilde{ {\cal H}},
\end{equation}
corresponding to the state given on the boundary of a disk with zero
area, and an operator
\begin{equation}
T: \tilde{\cal H}\otimes \tilde{\cal H}\rightarrow  \tilde{\cal H},
\end{equation}
corresponding to the intertwining operator associated with a pair of
pants (three-punctured sphere) of zero area.  By combining these objects
with the propagator described by the cylinder of nonzero area, it is
possible to build up an arbitrary Riemann surface.
Although the forms of $D$ and $T$ are trivial in the basis of
representations $| R \rangle$,  they have  a more complicated structure in
terms of the string basis.  For instance, the components of $D$ in the
string basis are given by
\begin{equation}
D_\sigma = \langle \sigma |\exp ( N\sum_{n \geq 1} a_n^{\dagger}) |0 \rangle,
\end{equation}
as was noted in \cite{douglas}.  In the string language, the state $D$
corresponds to the insertion of an $\Omega$-point singularity in the
set of coverings, where the covering maps can have a singularity
associated with an arbitrary permutation, and the operator $T$ can be
associated with an $\Omega^{-1}$-point type singularity (we will
review the description of these objects in the next section).  Again,
to calculate the partition function in the finite $N$ theory, one must
insert a projection operator at a single point on the manifold.
Although this approach gives a fairly simple understanding of the
finite $N$ result on a Riemann surface of general genus, we will not
elaborate on the details of this construction here; rather, we will
now derive the partition function for a Riemann surface of any genus
using the global group-theoretic techniques originally used in
\cite{gt1,gt2}, and verify that the result is in agreement with that
achieved by the Hilbert space approach in this section.

\section {Partition Function}
\setcounter{equation}{0}
\baselineskip 18.5pt

In \cite{gt1,gt2}, the expression (\ref{eq:partition}) for the
partition function was rewritten as an asymptotic series in $1/N$ by
expressing the coefficients in this expansion in terms of characters
of the symmetric group $S_n$, and then giving a geometric
interpretation in terms of the permutations of sheets of a covering
space realized by transport around nontrivial loops in the manifold
${\cal M}$.  The expansion of the partition function in $1/N$ was
accomplished by writing the dimension as well as the quadratic Casimir
of a representation associated with a Young diagram $R$ of $SU(N)$ in
terms of characters of the symmetric group representation associated
with the same Young diagram.  In this section, we show that using this
group-theoretic approach we can write the partition function for the
finite $N$ theory on a manifold of any genus, and that the projection
operator defined in Section 2 appears naturally as an object in the
algebra of the symmetric group.

It was shown in \cite{gt2} that using elementary formulae from the
theory of symmetric group representations, the contribution to the
partition function (\ref{eq:partition}) from all Young diagrams with a
finite number of boxes can be rewritten as
\begin{eqnarray}
Z_Y(G,\lambda A, N) &=&   \sum_{n}   \sum_{R \in Y_n} (\dim R)^{2 - 2G}
{\rm e}^{-\frac{\lambda A}{2 N}C_2(R)}  \nonumber\\
&=& \sum_{n, i, t,h\geq 0}
{\rm e}^{- \frac{n \lambda A}{2} }
\frac{(\lambda A)^{i + t + h}}{i!  \;t!  \; h!}
N^{n (2 - 2G)- i-2 (t + h)}
\frac{(-1)^i n^h (n^{2} - n)^t}{2^{t + h}}
\label{eq:result}\\
 & &\cdot \sum_{p_1, \ldots ,p_i \in T_2}
\sum_{s_1, t_1, \ldots ,s_G,t_G\in S_n} \left[
\sum_{R \in Y_n}
\frac{d_R}{n!^2}   \chi_R (p_1\cdots p_i \Omega_n^{2 - 2G}
\prod_{j =1}^{G}s_j t_j s_j^{-1} t_j^{-1}
)\right],
\nonumber \end{eqnarray}
where
\begin{equation}
\Omega_n = \sum_{\sigma \in S_n} \sigma N^{K_\sigma- n}
\label{eq:omega}
\end{equation}
is an element of the group algebra on $S_n$ with the property that
\begin{equation}
\dim R = {N^n\over n!} \chi_R(\Omega_n).
\label{eq:dimtwo}
\end{equation}

When discussing the large $N$ $SU(N)$ gauge theory, the contribution
{}from (\ref{eq:result}) is only part of the complete partition function.
In addition, one must consider representations corresponding to Young
diagrams with a finite number of columns with order $N$ boxes
(furthermore, it was shown by Douglas and Kazakov
\cite{dk} that for $G = 0$ and $\lambda A < \pi^2$ even this set of
representations is insufficient due to the phase transition at the point
$\lambda A = \pi^2$).  In the large $N$ case, (\ref{eq:result})
can be described as giving the contribution to the partition function
{}from a single chiral sector of the theory, corresponding to
orientation-preserving string maps from a world sheet onto ${\cal M}$.
The interpretation of this expression
in terms of string maps arises from taking the sum
over all representations $R$ of the symmetric group, giving
\begin{equation}
\delta (\rho) = \frac{1}{n!} \sum_{R}^{} d_R \chi_R (\rho),
\label{eq:identity2}
\end{equation}
where the $\delta$ function on the symmetric group algebra picks out
the coefficient of the identity permutation.
Inserting this result into (\ref{eq:result}), the contribution from
the group-theoretic terms reduces to
\begin{equation}
 \sum_{p_1, \ldots ,p_i \in T_2}
\sum_{s_1, t_1, \ldots ,s_G,t_G\in S_n}
N^{n(2-2 G)- i}
\left[\frac{1}{n!}   \delta (p_1\cdots p_i \Omega_n^{2 - 2G}
\prod_{j}s_j t_j s_j^{-1} t_j^{-1}
)\right].
\label{eq:case}
\end{equation}
This expression has a simple geometric interpretation as the sum of a
factor $N^{2-2 g}/| S_\nu |$ over all $n$-fold covers $\nu$ of ${\cal
M} $ by (possibly disconnected) surfaces of any genus $g$, where the
covering map $\nu$ has elementary branch points at a set of points
$w_1, \ldots, w_i$, and singularities associated with arbitrary
permutations of the covering sheets at a set of $2 -2G$ fixed points
called ``$\Omega$-points'' on ${\cal M}$ (if $G > 1$, there are $2G -
2$ ``$\Omega^{-1}$-points''; at these points there may be an arbitrary
number $x$ of $\Omega$-point type singularities all coalesced to a
single point, each carrying a factor of $-1$).  The geometric
interpretation of (\ref{eq:case}) can be derived by associating with
$s_j, t_j$ the permutations on sheets of a cover associated with
transport about a set of loops $a_j,b_j$ generating the first homotopy
group $\pi_1 ({\cal M})$, and similarly associating with $p_j$ the
permutations associated with transport around the elementary branch
points $w_j$.  Using this description of (\ref{eq:case}) in terms of
covering maps, it is possible to rewrite (\ref{eq:result}) in terms of
a sum of string maps.  We define $\Sigma ({\cal M})$ to be the set of
all orientation-preserving maps from a string world sheet to ${\cal
M}$ which are locally covering maps at all but a finite number of
singular points, where the singular points may be elementary branch
points, $\Omega$-points or $\Omega^{-1}$-points, or infinitesimal
tubes or handles contracted to points on ${\cal M}$.  We use again the
natural measure ${\rm d} \nu$ on $\Sigma ({\cal M})$ given by
associating with each singular point (other than the $\Omega$-points,
which are taken to be fixed) a measure factor proportional to the area
measure on ${\cal M}$ with proportionality constant $\lambda$.  In
terms of this set of string maps, the partition function in a single
chiral sector of the theory for large $N$ can be written as
\begin{equation}
Z_Y (G,\lambda A, N) = \int_{\Sigma ({\cal M})}  d \nu
\; {\rm e}^{- \frac{n \lambda A}{2} }
\frac{(-1)^{i} N^{2 -2 g}}{ |S_\nu |}
(-1)^{\sum_{j}x_j },
\label{eq:partitionsimple}
\end{equation}
where $n$ is the winding number of the map $ \nu$, $i$ is the number
of branch points in $\nu$, $| S_\nu |$ is the symmetry factor, $g$ is
the genus of the covering space, and $x_j$ are the numbers of
nontrivial twists at the $\max (0, 2G - 2) $ $\Omega^{-1}$-points in
the map $\nu$.

We will now study how this analysis must be modified when we have a
gauge group $SU(N)$ with finite $N$.  Restricting the sum over
representations to $Y^N_n$ for each $n$, the group theoretic terms in
expression (\ref{eq:result}) take the form
\begin{equation}
\sum_{p_1, \ldots ,p_i \in T_2}
\sum_{s_1, t_1, \ldots ,s_G,t_G\in S_n}
N^{n(2-2 G)- i} \sum_{R \in Y^{N}_n}
\left[\frac{d_R}{n!^2}
\chi_R (p_1\cdots p_i \Omega_n^{2 - 2G}
\prod_{j}s_j t_j s_j^{-1} t_j^{-1})\right].
\label{eq:finitecase}
\end{equation}

We now introduce a set of projection operators $P^{(N)}_n$.  For each
set of values $n, N$, $P^{(N)}_n$ is an element of the algebra over
the symmetric group $S_n$, defined by
\begin{equation}
P^{(N)}_n =\sum_{\sigma \in S_n}  \sum_{R \in Y^{N}_n}
\frac{d_R}{n!} \chi_R (\sigma) \sigma = \sum_{\sigma \in S_n}
N^{K_\sigma -n}P^{(N)} (\sigma) \sigma.
\label{eq:projection}
\end{equation}
These operators are projection operators in the sense that
\begin{equation}
(P^{(N)}_n)^2 = P^{(N)}_n,
\end{equation}
which follows directly from the relation
\begin{equation}
\frac{d_R}{n!}   \sum_{\sigma} \chi_R (\sigma)
\chi_S (\sigma^{-1} \tau) = \delta_{RS} \chi_R (\tau).
\end{equation}
Note that the projection operator $P^{(N)}_n$ commutes with all
elements of $S_n$.
By inserting the projection operator into a $\delta$  function on the
symmetric group, we have
\begin{equation}
\delta (\sigma P^{(N)}_n)  =
\sum_{R \in Y^{N}_n} \frac{d_R}{n!}  \chi_R (\sigma).
\label{eq:projectiond}
\end{equation}
Using this relation in the expression (\ref{eq:finitecase}) for the
group theoretic terms in the finite $N$ partition function, we get
\begin{equation}
\sum_{p_1, \ldots ,p_i \in T_2}
\sum_{s_1, t_1, \ldots ,s_G,t_G\in S_n}
N^{n(2-2 G)- i}
\left[\frac{1}{n!}
\delta (p_1\cdots p_i \Omega_n^{2 - 2G} P^{(N)}_n
\prod_{j}s_j t_j s_j^{-1} t_j^{-1})\right].
\label{eq:finited}
\end{equation}
Just as the expression (\ref{eq:case}) can be described in terms of a
sum over covering maps with $s_j,t_j$ representing the permutations on
the sheets of the cover given by transporting around the generators of
$\pi_1 ({\cal M})$, $p_j$ representing permutations around elementary
branch points, and the term $\Omega_n^{2 - 2G}$ giving the
contribution of $2 - 2G$ $\Omega$-points, we can interpret
(\ref{eq:finited}) in a similar fashion as a sum over covering maps.
All the possible singularities for covering maps in the large $N$ case
still appear here; in addition, however, there is a single fixed
``projection'' point on ${\cal M}$ where a singularity corresponding
to an arbitrary permutation can occur.  The weight associated with a
singularity giving a permutation $\sigma \in S_n$ is given by $P^{(N)}
(\sigma)$, just as we found in the Hamiltonian formulation.  Here,
this result arises by taking the expression from
(\ref{eq:projectiond}) and dividing by the factor $N^{K_\sigma - n}$
which appears because this singularity increases the genus of the
covering space.  We can now proceed to define a string theory of maps
for the finite $N$ theory for any genus $G$.  We define $\bar{\Sigma}
({\cal M})$ to be the set of all covering maps of ${\cal M}$ with a
finite number of the usual types of movable singularities (branch
points, infinitesimal tubes, and infinitesimal handles), $|2 - 2G |$
singularities of a general type at $\Omega$-points or
$\Omega^{-1}$-points, and an additional singularity at the projection
point associated with a permutation $\tau$ on the $n$ sheets of the
cover.  The partition function for finite $N$ of the QCD${}_2$ gauge
theory is then given by
\begin{equation}
Z (G,\lambda A, N) = \int_{\bar{\Sigma} ({\cal M})}  d \nu
\; {\rm e}^{- \frac{n \lambda A}{2} }
\frac{ N^{2 -2 g} }{ |S_\nu |}(-1)^{i+ \sum_{j} x_j}P^{(N)}(\tau).
\label{eq:finitepartition2}
\end{equation}

We have thus given a description of the partition function for finite
$N$ QCD${}_2$ on an arbitrary Riemann surface ${\cal M}$ in terms of a
weighted sum over a certain class of maps from a string world sheet
onto ${\cal M}$.  We have derived this result from two rather
different points of view and arrived at the conclusion that the
essential feature in the string picture for the partition function of
the finite $N$ theory is the insertion at a single point of a
singularity weighted with the coefficients of the projection operator
$P^{(N)}_n$.  Otherwise, the  string picture is extremely similar to
that in the large $N$ theory, with the change that the string
world sheet is not oriented in the finite $N$ theory.

A feature of this string description of the theory which complicates
the picture somewhat is the explicit appearance of factors of $N$, the
inverse string coupling, in the weights $P^{(N)} (\sigma)$ associated
with singularities at the projection point.  This dependence means
that the terms in the expansion cannot simply be ordered by genus
according to the power of $N$ they carry.  In the string picture, this
would correspond to string interaction terms with coefficients
proportional to the inverse string coupling; such terms make the
calculation of explicit results more complicated.  This also leads to
an interesting and somewhat nontrivial question about the relationship
between the finite $N$ and large $N$ theories.  Clearly, we expect
that the coefficients of the asymptotic $1/N$ expansion should really
be calculating the behavior of the finite $N$ theory as $N \rightarrow
\infty$.  From this string picture, however, this correspondence is
not obvious.  For example, in the large $N$ theory, when $G > 1$ we
expect the leading nontrivial
term in the  asymptotic expansion of the partition
function to be given by
\begin{equation}
Z_\infty (G, \lambda A, N) = 1 + 2 N^{2 - 2G} {\rm e}^{- \frac{\lambda
A}{2} } + {\cal O} (N^{2 - 2G - 2}).
\end{equation}
(We use the notation $Z_\infty$ to differentiate the asymptotic
expansion for large $N$ from the partition function at finite $N$.)
This implies that
\begin{equation}
\lim_{N \rightarrow \infty} N^{2G - 2}
\left( Z (G, \lambda A, N) - 1 \right) = 2 {\rm e}^{- \frac{\lambda
A}{2} }.
\label{eq:nontrivialimplication}
\end{equation}
Showing that (\ref{eq:nontrivialimplication}) is true from  the string
expression (\ref{eq:finitepartition2}) is a rather nontrivial
proposition.  In fact, once one has removed the $1/N$ ordering of the
asymptotic expansion for the partition function, it is difficult to
calculate anything concrete because in general an arbitrarily large
number of terms contribute to the partition function.  One might try
ordering in powers of ${\rm e}^{- \lambda A} $, but this is impossible
because arbitrarily large powers of $\lambda A$ appear.

\section {Conclusions}
\setcounter{equation}{0}
\baselineskip 18.5pt

We have seen that a string-theoretic description of the partition
function for finite $N$ QCD${}_2$ can be attained at the expense of
introducing an arbitrary ``projection point'' at which a permutation
of the sheets of the cover occurs.  Like the $\Omega$-points which
occur already in the large $N$ theory, the fact that these
singularities do not appear with area factors suggests that rather
than arising from a local interaction, they represent a global
phenomenon.  Indeed, this is rather clear in the case of the
projection point, since it is the expression of the fact that we may
obtain the finite $N$ theory by imposing the Mandelstam identities.
In string-theoretic terms, the Mandelstam identities may be regarded
as a {\it constraint} \cite{Baez}.  One complicating feature of the
finite $N$ string picture is that the weights associated with
singularities at the projection point contain explicit factors of $N$,
the inverse string coupling.  This implies that a Lagrangian
description of the theory on the world sheet will include string
interaction terms with coefficients proportional to the inverse string
coupling.

Recently it has been shown by Cordes, Moore, and Ramgoolam that the
counting of string maps with $\Omega$-point type singularities can be
expressed in a much simpler language by interpreting the sum over
$\Omega$-points as an orbifold Euler characteristic of a space of
covering maps, which appears naturally in the context of a
topological-type field theory on the world sheet \cite{cmr};
a similar world sheet topological theory was described in
\cite{horava} using harmonic maps.
It
would be interesting to explore the possibility of finding a similar
interpretation for the extra terms arising from the projection point
singularities; the fact that these singularities carry factors of $N$
rather then $1/N$, however, may complicate such an
interpretation.

An interesting challenge will be to calculate the vacuum expectation
values of Wilson loops for finite $N$ QCD${}_2$ in the
string-theoretic framework.  From the results for Wilson loops in the
large $N$ theory \cite{gt2} we expect the result to be expressed in
terms of a sum over maps from open string world sheets with boundaries
which are taken to the Wilson loops by the string maps.  For $SU(2)$,
it seems that such a description of Wilson loop VEV's should be
possible.  In this case, the Wilson loops are not oriented, and a
Wilson loop can thus bound a string world sheet on either side.
Mathematically, this is simply a result of the fact that the
fundamental and the conjugate representations are identical for
$SU(2)$.  Even for $SU(2)$, a rigorous description of Wilson loop
VEV's in terms of a sum over string maps is complicated by
self-intersections of the Wilson loops.  In principle, one should be
able to construct such a description with self-intersections described
by twists in the string world sheet, as was done for the large $N$
theory in \cite{gt2}.  It would be interesting to work through the
details of this calculation and compare to results on Wilson loops in
the $SU(2)$ theory achieved through other methods, such as in
\cite{ek}.  For finite values of $N$ greater than 2, the problem is
complicated further by the fact that the tensor product of $N$
fundamental representations contains the trivial representation.
Thus, it is difficult to see how to bound a string world sheet by
single Wilson loops.  Physically, this obstruction corresponds to the
existence of baryons, which are nonexistent in the large $N$ theory,
and which are essentially equivalent to mesons for $SU(2)$.  Finding a
clean string-theoretic description of Wilson loop VEV's for $N > 2$
may require an alternative to the ``projection point'' formalism we
have described here.

Finding a string interpretation of Wilson loop VEV's for finite values
of $N$ would, however, shed interesting new light on theories other
than QCD${}_2$, as described in \cite{Baez}.  First, if one could
simply handle the $\lambda \to 0$ limit of the theory one could obtain
a string-theoretic interpretation of $BF$ theory, also known as
topological Yang-Mills theory.  (For a review of this theory, see for
example \cite{BlauT}.)  When the gauge group is $SO(3)$, the Wilson
loop vacuum expectation value $\langle W(\gamma_1) \cdots W(\gamma_n)
W(\eta_1) \cdots W(\eta_m)\rangle$ in $BF$ theory may also be
interpreted as the inner product of string states $\langle \gamma_1,
\dots, \gamma_n|\eta_1, \dots, \eta_m\rangle$ in Euclidean quantum
gravity on the manifold ${\bf R} \times {\cal M}$.  In the quantum
gravity context, one may think of the curves $\eta_i$ as
living in $\{t_1\}
\times {\cal M}$, the curves $\gamma_i$ as living in $\{t_2\} \times
{\cal M}$, and the string world sheets as being mapped into $[t_1,t_2]
\times {\cal M}$.  (Since the Hamiltonian constraint in
quantum gravity effectively makes the Hamiltonian zero, the results
are independent of $t_1$ and $t_2$.)  These expectation values can
then be interpreted as describing an inner product on the physical
Hilbert space of gravity in the loop variable formulation.  Such a
construction would give a natural interpretation of Euclidean quantum
gravity in 3 dimensions as a string theory.

\vskip .5truein
{\Large{\bf Acknowledgements}}

W.\ T.\ would like to thank O.\ Alvarez and D.\ Gross for helpful
discussions.

\end{document}